\documentclass[aps,pra,twocolumn,groupedaddress,amsmath,amssymb,10pt]{revtex4}
\usepackage{epsfig}
\usepackage{amssymb}
\usepackage{amsmath}
\usepackage{graphicx}
\usepackage{amsfonts}
\usepackage{epsfig}
\usepackage{revsymb}
\usepackage{bm}
\usepackage{amsmath}
\usepackage{amssymb}
\usepackage{latexsym}
\usepackage{hyperref}
\usepackage{cleveref}

\begin{document}

\title{Revisiting the spin-orbit scattering 
in small-sized superconducting particles \\ in the magnetic field}
\author{Serguei N. Burmistrov}
\begin{abstract}
The Knight shift of nuclear magnetic resonance is an experimental probe of the paramagnetic spin susceptibility in metals. Information about the electron pairing in superconductors can be extracted from the Knight shift in the small-sized particles. The finite zero-temperature magnitude of paramagnetic susceptibility observed in the superconducting particles has been associated with the spin-orbit scattering of conduction electrons. The conventional treatment has delivered the paramagnetic susceptibility magnitude proportional to the square of the  spin-orbit interaction amplitude. Here we examine the coupling between the superconducting current and the spin-orbit scattering of conduction electrons in the small-sized particles. Such interference coupling, absent in the normal state, results in an additional spin polarization of the conduction electrons generating the superconducting current in the magnetic field. This anomalous effect delivers the noticeably larger contribution to the paramagnetic susceptibility. The magnitude of the effect as well as the contribution to the paramagnetic susceptibility prove to be proportional to the amplitude of spin-orbit interaction. The frequency of electron paramagnetic resonance is shifted as well.  
\end{abstract}
\maketitle

\section{Introduction}
\par 
The Knight shift \cite{Knight} implies a shift of the nuclear magnetic resonance frequency for the atoms in a metal as compared to the same atoms in dielectrics. The shift originates from an additional local magnetic field produced with the Pauli paramagnetic spin susceptibility of conduction electrons. The magnitude of the frequency shift \cite{Reif,Knight1} is proportional to the paramagnetic susceptibility $\chi$ and reads 
$$
\Delta\Omega /\Omega= \bigl(8\pi n_{\text{at}}/3\bigr)\vert\psi (0)\vert ^2\chi 
$$
where $n_{\text{at}}$ is the number of atoms per unit volume and $\vert\psi (0)\vert ^2$ is the probability density to find an electron at the position of the nucleus. 
\par 
According to the BCS theory, the paramagnetic susceptibility of a superconductor should completely  vanish at zero temperature \cite{Yosida}. To explain a finite magnitude of the Knight shift observed at low temperatures in the conventional superconductors with the spin-singlet pairing, Ferrel \cite{Ferrel} and Anderson \cite{Anderson} have put forward an assumption about  a key role of spin-orbit interaction between electrons in small-sized superconducting particles. Owing to spin-orbit scattering, the electron can change its spin direction, entailing nonzero paramagnetic susceptibility of a superconductor at zero temperature \cite{Abrikosov}. In fact, in this case the electron states can no longer be characterized by the eigenvalues of the electron spin. 
\par 
In Refs. \cite{Maki,Larkin} there has been studied the effect of magnetic field close to the critical one on the magnitude of the Knight shift in the superconducting particles of the sizes smaller than the magnetic field penetration depth and the coherence length. However, in the above-cited works a possible interplay  between the scattering on the spin-independent and spin-orbit parts of impurity-electron interaction has completely been disregarded. 
\par 
Recently \cite{Fukaya}, the paramagnetic spin susceptibility for the orbital-singlet Cooper pairing has been studied on the example  of superconducting Sr$_2$RuO$_4$. As is expected, the paramagnetic spin susceptibility  reduces below the superconducting transition with decreasing the temperature. However,  owing to strong atomic spin-orbit coupling \cite{Puetter},  the paramagnetic  spin susceptibility does not vanish at zero temperature and conserves its noticeably finite magnitude at zero temperature. The behavior of the Knight shift is observed to be unusual \cite{Saitou} in a bulk organic superconductor with quasi-2D triangular lattice. The paramagnetic spin susceptibility retains the normal state magnitude even deep in the superconducting state.  
\par 
In superconductors \cite{Levitov} there may appear an additional spin polarization induced with the superconducting current in the process of spin-orbit scattering of electrons with the impurities. In particular, the coupling between the spin-orbit scattering and the superconducting current in the magnetic field should result in the finite magnitude of paramagnetic susceptibility at zero temperature. In some sense this mechanism is analogous both to the spin polarization with the electric current in a normal metal \cite{Dyakonov} and to the spin transport in nanoscopic structures \cite{Eto,Reynoso} and irregular-shaped samples \cite{Chiang} with the spin-orbit Rashba interaction. The phenomenon is often referred to as the spin Hall effect \cite{Hirsch}. The magnetization induced with the electric current is found in tellurium \cite{Furukawa} and in compound UNi$_4$B \cite{Saito}. This magnetoelectric effect \cite{Mineev} is associated with the spin-orbit coupling of electrons in the non-centrosymmetric crystals where the spin and orbital  degrees of freedom can be mixed. 
\par 
The spin-orbit coupling is essential for the properties of  proximity-induced superconductivity in a normal metal attached to a superconductor \cite{Mishra} and for two-dimensional 
superconductor/magnet heterostructures \cite{Bobkov}. These objects can be promising for the superconducting spintronics.  
\par 
In addition, the spin-orbit scattering in the magnetic field should lead both to the inhomogeneous energy  dispersion in the density of states for the electrons of opposite spin directions and to the shift of electron paramagnetic resonance frequency. 
 \par 
 Here we treat the interference between the spin-independent  and spin-orbit scatterings with an impurity in the small-sized superconducting particles in the magnetic field. Such interference effect should noticeably enhance the zero-temperature paramagnetic susceptibility and change the frequency of the electron paramagnetic resonance.  

\section{Effective Bohr magnetons}

\par 
To describe the electron spin-flip effects in the small-sized superconducting particles, we employ the approach suggested in Ref.~\cite{Abrikosov}. In the mathematical formulation this approach reduces to  the problem on the scattering of an electron with the randomly localized impurities  having the spin-orbit component. The scattering amplitude $f(\bm{p},\bm{p}')$ can be written as 
\begin{equation}
\label{eq1}
f(\bm{p},\bm{p}')=a(\bm{p},\bm{p}')+ib(\bm{p},\bm{p}')\frac{\bm{\sigma} (\bm{p}\times\bm{p}')}{p_F^2}
\end{equation}
where $\bm{p}$ and $\bm{p}'$ are the momenta of an electron before and after the collision, $\bm{\sigma}$ implies the Pauli matrices, and $p_F$ is the Fermi momentum. The amplitude $b(\bm{p},\bm{p}')$ describes the spin-orbit interaction and is commonly of the order $\vert b/a\vert\sim Z\alpha$. Here, as usual, $\alpha =e^2/\hbar c$ is the fine-structure constant and $Z$ is the atomic number of an impurity. 
\par 
To have the calculation simple and obvious, we keep inequality $\vert b/a\vert\ll 1$ in mind. In other words, we assume that the mean free path $l_s$ with respect to the spin-orbit scattering is much larger as compared with the mean free path $l$ between the collisions of electrons with the impurities. The latter means that the free path of an electron is mainly determined by amplitude $a(\bm{p},\bm{p}')$ in 
Eq.~\eqref{eq1}. 
\par 
So, we put $l_s\gg l$ and neglect the terms having the smallness $\vert b/a\vert^2$ and higher in our consideration. In what follows, we take the scattering amplitudes $a(\bm{p},\bm{p}')$ and $b(\bm{p},\bm{p}')$ to be isotropic and suppose the dirty limit as $l_s\gg\xi_0\gg l$ where $\xi_0$ is the coherence length of a superconductor. We imply hereafter $\hbar=1$ and the size of  small superconducting particles less than the penetration depth in the magnetic field. 
\par 
For the small-sized superconductors, we find three types of contributions to the normal and anomalous  Green functions in the linear approximation in the  uniform magnetic field $\bm{H}$. The first type originates from the diamagnetic superconducting current.  The second is regular and due to splitting the magnetic energy $\mu_0\bm{\sigma}\bm{H}$ for the different spin projections, $\mu_0$ being the Bohr magneton. The third, most interesting and unexpected, comes from the interference of the terms $a(\bm{p},\bm{p}')$ and $b(\bm{p},\bm{p}')$ in the scattering amplitude $f(\bm{p},\bm{p}')$ \eqref{eq1} with the superconducting current.  These contributions can be represented as  $\mu_i\bm{\sigma}\bm{H}$ with some effective Bohr magnetons $\mu_i=\mu_i(\varepsilon)$ ($i=1, 2, 3,4)$ depending on the excitation energy.  In general, four kinds of effective magnetons $\mu_i(\varepsilon)$ are expected as a result of four possibilities to obtain the cross interference terms from two various  scattering amplitudes $a(\bm{p},\bm{p}')$ and $b(\bm{p},\bm{p}')$. As we will see below, the four possible effective magnetons, in essence,  are reduced to two different types $\mu_1(\varepsilon)$ and $\mu_2(\varepsilon)$. Such interference of the amplitudes $a(\bm{p},\bm{p}')$ and $b(\bm{p},\bm{p}')$ is absent in the normal state of a superconductor.  
\par
\begin{figure}[ht]
\begin{center}
\includegraphics[scale=0.6]{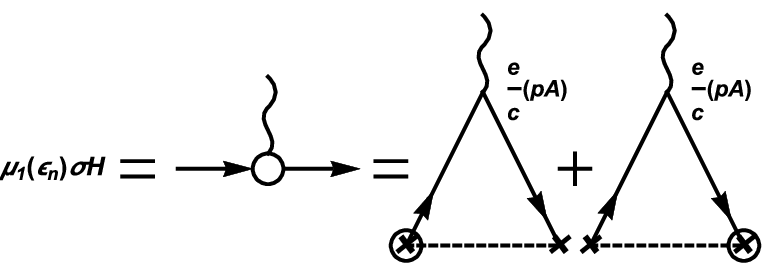}
\caption{The cross $\times$ means the spin-independent amplitude $a(\bm{p},\bm{p}')$ and the circled cross $\bigotimes$ denotes the spin-orbit amplitude $b(\bm{p},\bm{p}')$. The other terms corresponding to the four possible directions of arrowheads at the diamagnetic vertex $(e/c)\bm{pA}$ are not shown for brevity. The solid lines imply either normal or anomalous Green functions of a superconductor. The dashed line signifies the scattering  of electron at the same impurity. }
\label{BohrMag1}
\end{center}
\end{figure}

\begin{figure}[ht]
\begin{center}
\includegraphics[scale=0.56]{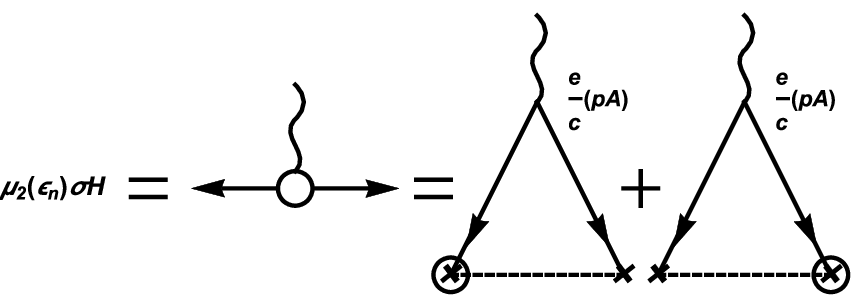}
\caption{The cross $\times$ means the spin-independent amplitude $a(\bm{p},\bm{p}')$ and the circled  cross $\bigotimes$ denotes the spin-orbit amplitude $b(\bm{p},\bm{p}')$. The other terms corresponding to the four possible directions of arrowheads at the diamagnetic vertex $(e/c)\bm{pA}$ are not shown for brevity. The solid lines imply either normal or anomalous Green functions of a superconductor. The dashed line signifies the scattering of electron at the same impurity.  }
\label{BohrMag2}
\end{center}
\end{figure}

 
The first effective Bohr magneton $\mu_1\bm{\sigma}\bm{H}$ is represented in Fig.~\ref{BohrMag1}. 
The second effective Bohr magneton $\mu_2\bm{\sigma}\bm{H}$ differs from $\mu_1\bm{\sigma}\bm{H}$ with the direction of the exit arrows alone and is given in Fig.~\ref{BohrMag2}. The other two remaining magnetons are reduced to these two ones $\mu_1(\epsilon)$ and $\mu_2(\epsilon)$ considered. In essence, there appear only two different magnetons which have the arrowheads in the same direction or in the opposite one. 
\par 
Let us give the expressions related to the two various  effective Bohr magnetons depending on the  Matsubara odd frequency $\varepsilon_n=\pi T (2n+1)$ ($n=0, \pm 1, \pm 2\dots$):
\begin{gather*}
\mu_1(\varepsilon _n)=\mu_0\frac{\beta}{2\tau\eta(\varepsilon_n)}\frac{\Delta^2}{\bigr(\Delta^2+\varepsilon_n^2\bigr)^{3/2}},
\\
\mu_2(\varepsilon _n)=\mu_0\frac{\beta}{2\tau\eta(\varepsilon_n)}\frac{i\varepsilon_n\Delta}{\bigl(\Delta^2+\varepsilon_n^2\bigr)^{3/2}}.
\end{gather*}
As usual,  $\Delta$ is the superconducting gap and we have denoted 
\begin{gather*}
\eta (\varepsilon_n)=1+\frac{1}{2\tau\sqrt{\Delta^2+\varepsilon_n^2}}, 
\\
\frac{1}{\tau}=\frac{nmp_F}{4\pi^2}\int \vert a\vert^2 d\Omega\quad\text{and}\quad  
\beta =\frac{2}{3}\sqrt{\frac{l}{l_s}}
\end{gather*}
where $\tau$ is the mean free time expressed via impurity concentration $n$ and electron mass $m$.  
We define the mean free time $\tau_s$ with respect to the spin-orbit scattering according to
$$
\frac{1}{\tau_s}=\frac{nmp_F}{4\pi^2}\int \vert b\vert^2\sin^2\theta\, d\Omega.
$$
The corresponding mean free paths $l$ and $l_s$ are labeled as $l=v_F\tau$ and $l_s=v_F\tau_s$, $v_F$ being the Fermi velocity. We note here that such interference is inherent in the superconducting state alone and vanishes if superconducting parameter $\Delta =0$.
\par 
The magnitude 
\begin{equation}
\beta =\frac{\chi_s (0)}{\chi_n}=\frac{2}{3}\sqrt{\frac{l}{l_s}},
\label{parsus}
\end{equation}
as we will see below, represents a ratio of paramagnetic spin susceptibility $\chi_s(0)$ at zero temperature to its magnitude $\chi_n$ in the normal state. We emphasize here that the interference between the spin-independent and spin-orbit  scatterings in the magnetic field results in the contribution to the paramagnetic susceptibility $\chi_s(0)$ proportional to $1/\sqrt{l_s}\sim Z\alpha$. The contribution from the spin-orbit scattering alone \cite{Abrikosov,Larkin} gives the weaker behavior in the large $l_s\gg l$ limit as $\chi_s(0)\sim 1/l_s\sim (Z\alpha)^2$.  
\par
\begin{figure*}[ht]
\begin{center}
\includegraphics[scale=0.8]{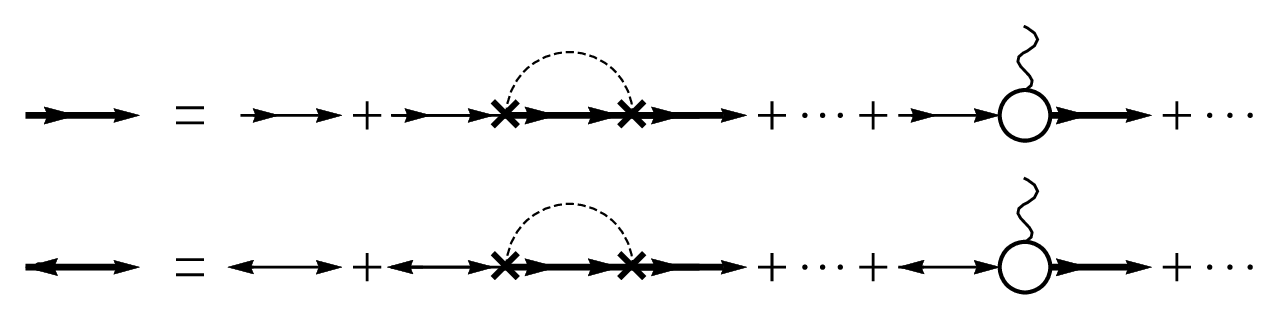}
\caption{The self-consistent equations for the normal and anomalous Green functions are shown diagrammatically. The thin lines illustrate the bare Green functions and the thick ones correspond to the renormalized Green functions after averaging over the impurity sites distributed chaotically in the particle. The cross $\bm{\times}$ denotes the scattering with the nonmagnetic part of impurity. The dashed semicircle line means the scattering at the same impurity. The circle and wavy line imply the effective Bohr magneton and interaction $\mu_{\text{eff}}\bm{\sigma}\bm{H}$. The similar and regular interaction $\mu_0\bm{\sigma}\bm{H}$ is not shown. The diagrams with the other possible directions of arrows for the Green functions are omitted for shortness.}
\label{BohrMag3}
\end{center}
\end{figure*}

 
In order to find the spectrum of one-particle excitations, one should know the total normal $\frak{G}$ and anomalous $\frak{F}^+$ Green functions which we determine from the system of diagrams shown in Fig.~\ref{BohrMag3}. In Fig.~\ref{BohrMag3} all the possible directions of arrows are omitted in the self-energies. The term with the regular interaction $\mu_0\bm{\sigma}\bm{H}$, entailing the simple shift in energy, will be restored in the final answer. 
\vspace*{65pt}
\par 
However, such procedure might drive us to the conclusion about instability of superconducting state in any 
small magnetic field. This can be associated with the specific features for the excitation spectrum in a superconductor due to noticeable magnitude of effective Bohr magnetons $\mu_1, \,\mu_2\sim\mu_0\beta\Delta^2/\sqrt{\Delta^2-\varepsilon^2}$ near energy $\varepsilon\sim\Delta$. The physical situation is the following. Due to such singularity the energy connected  with the effective Bohr magnetons would exceed the binding energy of an electron Cooper pair. This should result in breaking the superconducting state down. 
\par 
To inspect such suspicion, we return to the diagrams shown in Fig.~(\ref{BohrMag3}). It is necessary to involve the contribution from the diamagnetic superconducting current, resulting in smoothing the singularities in the spectrum of states and effective Bohr magnetons. On the whole, this leads to restoring and conserving the superconducting state. 
\par 
The behavior of small-sized superconducting particles in the constant magnetic field is studied in Refs.~\cite{Larkin,Maki}. We below restrict ourselves with the case of dirty limit $\tau\Delta\ll 1$ and $l\ll d$ where $d$ is the size of particles. 
\par 
The normal $\frak{G}$ and anomalous $\frak{F}^{+}$ Matsubara Green functions of a superconductor in the magnetic field are attempted  within the framework of the Gor'kov equations which describe perfectly the microscopic BCS superconductivity. So, we write a set of coupled equations on the Matsubara $\frak{G}$ and $\frak{F}^+$ Green functions
\begin{gather}
\biggl[ix-\frac{(\bm{p}-e\bm{A}/c)^2}{2m}+\mu\biggr]\frak{G}+y\frak{F}^{+}=\hat{1}, \nonumber
\\
\biggl[ix +\frac{(\bm{p}+e\bm{A}/c)^2}{2m}-\mu\biggr]\frak{F}^{+}+y\frak{G}=\hat{0} 
\label{GF}
\end{gather}
where variables $ix$ and $y$ are the self-energies to be determined. 
\par 
Restricting the terms in the magnetic field not higher than the quadratic ones,  we  obtain the following relations corresponding to the diagrams in  Fig.~(\ref{BohrMag3}): 
\widetext
\begin{gather*}
ix=i\varepsilon_n+\frac{1}{2\tau}\frac{ix}{\sqrt{y^2+x^2}}+\frac{\gamma}{2\tau}\frac{ix}{\sqrt{y^2+x^2}}\frac{2y^2-x^2}{(y^2+x^2)^2}-\frac{\beta\mu_0\bm{\sigma}{\bm{H}}}{2\tau}\frac{y^2}{(y^2+x^2)^{3/2}}, 
\\
y=\Delta +\frac{1}{2\tau}\frac{y}{\sqrt{y^2+x^2}}+\frac{\gamma}{2\tau}\frac{y}{\sqrt{y^2+x^2}}\frac{y^2-2x^2}{(y^2+x^2)^2}-\frac{\beta\mu_0\bm{\sigma}{\bm{H}}}{2\tau}\frac{ixy}{(y^2+x^2)^{3/2}}.
\end{gather*}
\widetext
\twocolumngrid 
\noindent
Here we have introduced 
$$
\gamma =\frac{1}{3}\frac{e^2}{c^2}v_F^2\langle \bm{A}^2\rangle_V\sim (\mu_0H)^2(p_Fd)^2
$$
where $\langle \bm{A}^2\rangle_V$ represents the average  of the vector-potential square over the volume  $V$ of superconducting particle.  
Then we put 
\begin{equation}\label{zet}
\zeta=2\tau\gamma=\frac{2}{3}\frac{e^2}{c^2}v_F^2\tau\langle\bm{A}^2\rangle_V
\end{equation}
and $u=ix/y$. After the algebraic transformations we obtain, returning the regular shift 
$\mu_0\bm{\sigma}\bm{H}$ in energy, that 
$$
ix =u\eta (u)\quad\text{and}\quad y=\Delta \eta (u)  
$$
where variable $u$ satisfies the following equation:
\begin{equation}\label{eq6}
u\biggl(1-\frac{\zeta}{\sqrt{\Delta^2-u^2}}\biggr)=i\varepsilon_n+\mu_0\bm{\sigma}\bm{H}
-\frac{\beta\mu_0\bm{\sigma}\bm{H}}{1+2\tau\sqrt{\Delta^2-u^2}}. 
\end{equation}
\par
The spectral density of one-particle excitations is determined by the relation 
\begin{equation}\label{eq7}
\rho (\varepsilon)=\rho_n \,\text{Im}\, \frac{u}{\sqrt{\Delta^2-u^2}}
\end{equation}
where variable $u$ satisfies equation \eqref{eq6} continued analytically from the upper half-plane. The spectral density $\rho _n$ corresponds to that in the normal state. The energy gap $\varepsilon_0$ in the spectrum  according to equation  \eqref{eq6} reads 
$$
\varepsilon_0=2\Delta \bigl(1-\zeta^{2/3}\bigr)^{3/2}-2(1-\beta)\mu_0 H.
$$
From Eqs. \eqref{eq6} and \eqref{eq7} one can see that there occurs an inhomogeneous splitting for the spectrum of states with the different spin projections by quantity $2(1-\beta)\mu_0H$ at $\tau\varepsilon\ll 1$ and by usual quantity $2\mu_0H$ at $\tau\epsilon\gg 1$. 

\section{Electron paramagnetic resonance}

\par 
The electron paramagnetic resonance frequency is determined by the transitions between the states with the different spin projections. The main contribution is delivered with the excitations having the energy of about temperature. Thus, for $\tau T\ll 1$, the resonance frequency will be $\Omega_{res}=2(1-\beta)\mu_0H$. If $\tau T\gg 1$, there is no additional shift and the resonance frequency remains unchanged, i.e. $\Omega _{res}=2\mu_0H$. The absorption will have the width equal to $2\beta\mu_0H$ since only the splitting of the density of states is significant in the field $H\rightarrow 0$.  As a result, one can neglect the square of the magnetic field proportional to $\zeta$ in Eq. \eqref{zet} and approximate the normal 
$\frak{G}$ and anomalous $\frak{F}$ Matsubara Green functions according to  
\begin{gather}\nonumber
\frak{G}(\varepsilon_n)=-\frac{\xi +u\eta (u)}{\xi^2 +(\Delta^2-u^2)\eta^2(u)},
\\
\frak{F}(\varepsilon_n)=\frac{\Delta\eta (u)}{\xi^2 +(\Delta^2-u^2)\eta^2(u)}, \label{eq8}
\\
u=i\varepsilon_n+\mu_0H-\frac{\beta\mu_0H}{1+2\tau\sqrt{\Delta^2-u^2}}. \nonumber
\end{gather}
Here, as usual, $\xi =(\bm{p}^2/2m -\mu)$ is the electron energy taken from the Fermi energy $\mu$. 
\par 
Next, we use the routine expression for the transverse paramagnetic susceptibility 
\begin{multline*}
\chi (\omega_n)=-2\mu_0^2T\sum\limits_{\varepsilon_m}\int\frac{d^3p}{(2\pi)^3}\times
\\ 
\biggl[\frak{G}_{-}(\varepsilon_m +\omega_n/2)\frak{G}_{+}(\varepsilon_m-\omega_n/2) 
\\
+\frak{F}_{-}(\varepsilon_m +\omega_n/2)\frak{F}_{+}^+(\varepsilon_m-\omega_n/2)\biggr].
\end{multline*}
Here the symbols $\frak{G}_{\pm}$ and $\frak{F}_{\pm}$ are the Matsubara Green functions in the magnetic field and differ with the sign of magnetic field in Eq.~\eqref{eq8}.  The elastic scattering with the static impurities does not change the energy of an electron. Then, after averaging the product of two Green functions over the sites of randomly localized impurities and integrating over the momentum, the same expressions appear as the ones for the  pure superconductor without factors $\eta (u)$. 
\par 
Performing the analytical continuation in the Matsubara frequency $\omega _n$ from the upper complex  half-plane to the real frequency $\omega$, we arrive at  the rather complicated expression where $G^{R(A)}$ and $F^{R(A)}$ are the retarded and advanced Green functions of a superconductor:
\widetext
\begin{multline}\label{eq10}
\chi (\omega)=-2\mu_0^2\int\frac{d\varepsilon}{4\pi i}\int\frac{d^3p}{(2\pi)^3}\biggl[\biggl(G_{-}^R\bigl(\varepsilon +\frac{\omega}{2}\bigr)G_{+}^R\bigl(\varepsilon -\frac{\omega}{2}\bigr)+ F_{-}^R\bigl(\varepsilon +\frac{\omega}{2}\bigr)F_{+}^{+R}\bigl(\varepsilon -\frac{\omega}{2}\bigr)\biggr)\tanh\frac{\varepsilon -\omega/2}{2T}
\\
- \biggl(G_{-}^A\bigl(\varepsilon +\frac{\omega}{2}\bigr)G_{+}^A\bigl(\varepsilon -\frac{\omega}{2}\bigr)+ F_{-}^A\bigl(\varepsilon +\frac{\omega}{2}\bigr)F_{+}^{+A}\bigl(\varepsilon -\frac{\omega}{2}\bigr)\biggr)\tanh\frac{\varepsilon +\omega/2}{2T}
\\ 
+\biggl(G_{-}^R\bigl(\varepsilon +\frac{\omega}{2}\bigr)G_{+}^A\bigl(\varepsilon -\frac{\omega}{2}\bigr)+ F_{-}^R\bigl(\varepsilon +\frac{\omega}{2}\bigr)F_{+}^{+A}\bigl(\varepsilon -\frac{\omega}{2}\bigr)\biggr)\biggl(\tanh\frac{\varepsilon +\omega/2}{2T}-\tanh\frac{\varepsilon -\omega/2}{2T}\biggl)\biggr].
\end{multline}
\widetext
\twocolumngrid
Inserting the retarded $G^R (F^R)$ and advanced $G^A (F^A)$ Green functions continued analytically from the corresponding half-planes into Eq. \eqref{eq10}, we  represent  the frequency behavior of the transverse susceptibility $\chi (\omega)$ in the vicinity of resonance frequency as 
\begin{multline*}
\frac{\chi(\omega)}{\chi_n}=-\int_{\Delta}^{\infty}\frac{d\varepsilon}{2T}\frac{\varepsilon}{\sqrt{\varepsilon^2-\Delta^2}}\frac{1}{\cosh^2(\varepsilon /2T)}
\\ 
\times\frac{\Omega_0}{\omega -\Omega_0+\frac{\beta\Omega_0}{1+4\tau^2(\varepsilon^2-\Delta^2)}+i\delta}.
\end{multline*}
Here we denote $\Omega_0=2\mu_0H$ as a conventional electron paramagnetic resonance frequency in the magnetic field $H$. The electron paramagnetic resonance  frequency $\Omega_{res}(T)$ is determined from condition $\text{Re}\,\chi(\Omega_{res})=0$. 

\section{Conclusion}
\par 
In summary, we have examined  the coupling of the superconducting current with the spin-orbit scattering of conduction electrons in the small-sized particles of a superconductor in the magnetic field. 
To simplify the physical description, the particle size is assumed to be no larger than the penetration depth of the magnetic field. The interplay of superconducting current with the spin-orbit part of the electron scattering can be described in the terms of some effective Bohr magnetons $\mu_1(\epsilon)$ and $\mu_2(\epsilon)$ additional to the usual electron Bohr magneton $\mu_0$. One effective Bohr magneton $\mu_1(\varepsilon)$ is associated with the normal Green function of a superconductor. The other 
$\mu_2(\varepsilon)$ is done with the anomalous Green function. The both effective Bohr magnetons 
 are proportional to the amplitude of spin-orbit interaction and vanish in the normal state when the superconducting gap $\Delta =0$. 
\par 
The interplay of superconducting current with the spin-orbit scattering leads to an additional spin polarization. In some sense this is similar to the spin Hall effect. As is compared with the conventional mechanisms, the effective Bohr magnetons deliver a noticeable contribution to the paramagnetic susceptibility. The paramagnetic susceptibility in the small-sized superconducting particles remains finite at zero temperature and proportional to the inverse square root from the mean free path $l_s$ with respect to  the spin-orbit scattering of electrons. 
\par
In the magnetic field the spin-orbit scattering  results both in the inhomogeneous energy  dispersion in the density of states for the electrons with the opposite spin directions and in the shift of electron paramagnetic resonance frequency.

\end{document}